\documentstyle[twocolumn,aps,prl,epsf,floats]{revtex}

\begin{document}
\draft
%
%
\wideabs{
%
%
\title{Optical studies of charge dynamics in the optimally-doped \boldmath
Bi$_2$Sr$_2$CaCu$_2$O$_{8+\delta}$ and the non-superconducting
(Bi$_{0.5}$Pb$_{0.5}$)$_2$Ba$_3$Co$_2$O$_\delta$ single crystals \unboldmath}

%
%
\author{J.J.~Tu$^1$, C.C.~Homes$^1$, G.D.~Gu$^1$, D.N. Basov$^2$, S.M.
Loureiro$^3$, R.J. Cava$^3$, and M. Strongin$^{1,}$\cite{correspond} }
%
%
\address{$^1$Department of Physics, Brookhaven National Laboratory, Upton,
New York 11973}
\address{$^2$Physics Department, University of California at San Diego, La
Jolla, California 92093-0319}
\address{$^3$Department of Chemistry and Materials Institute, Princeton
University, Princeton, New Jersey 92093-0319}

\date{\today}
\maketitle
%
%
\begin{abstract}
An analysis of the frequency-dependent scattering rate $\tau^{-1}(\omega)$
reveals signatures of the superconducting gap as well as the pseudogap in
high-$T_c$ cuprates.  These features can be identified in the
temperature-dependent spectral function $W(\omega)$, inverted from the optical
data of the optimally-doped Bi$_2$Sr$_2$CaCu$_2$O$_{8+\delta}$ (Bi2212) using
an experimentally unambiguous method that shows the behavior of both the maxima
and minima in the spectral function.
\end{abstract}
\pacs{PACS: 74.25.Gz, 74.72.Hs, 74.25.Nf}
%
%
}
%
%
%
\narrowtext%
\vspace*{-0.8cm}
%
%
For superconductors of the type described by the Bardeen-Cooper-Schrieffer
(BCS) theory, the spectral function $\alpha^2F(\omega)$ contains all the
relevant informa-tion con\-cerning superconductivity \cite{carbotte00}.
Experimentally, $\alpha^2F(\omega)$ is commonly obtained from tunneling I-V
curves.  These measurements reveal features in the quasiparticle density of
states that are the results of electron-phonon interaction and with the well
developed Eliashberg theory can lead to the determination of
$\alpha^2F(\omega)$ \cite{mcmillian65}.  In principle the optical conductivity
and the frequency-dependent scattering rate can also be used to determine the
spectral function, $\alpha^2F(\omega)$, for BCS superconductors
\cite{allen71,farnworth76,marsiglio98}.  Recently it has been recognized the
same kind of analysis can be used to determine a spectral function $W(\omega)$
for {\it d}-wave high-$T_c$ superconductors \cite{carbotte99} from the optical
data.  This has lead to the conclusion that charge coupling to the spin
resonance \cite{abanov01} found in neutron scattering is the analogue to the
common electron-phonon interaction in BCS superconductors.  We will, however,
avoid addressing directly the question of whether high-$T_c$ cuprates are BCS
superconductors. Instead, we will take a more empirical approach to discuss the
common features in the spectral functions of different high-$T_c$ cuprates and
how they move with temperature, especially around $T_c$.

In this Letter, we report an analysis of the temperature-dependent spectral
function $W(\omega)$, obtained from an inversion of new optical data for
optimally-doped Bi2212 using an experimentally unambiguous technique that
examines the maxima and minima in $W(\omega)$ simultaneously.  A comparison with
the spectral function of optimally-doped YBa$_2$Cu$_3$O$_{6.95}$ (YBCO) suggests that a
pseudogap exists in optimally-doped Bi2212 above $T_c$ at 100~K, but not in
optimally-doped YBCO and that the pseudogap can co-exist with the
superconducting gap in optimally-doped Bi2212 below $T_c$.  The spectral
function of the metallic non-superconducting (non-SC) cobaltate
(Bi$_{0.5}$Pb$_{0.5}$)$_2$Ba$_3$Co$_2$O$_\delta$ single crystals, which are
isomorphic to Bi2212 in terms of crystal structure
\cite{watanabe91,terasaki93,loureiro01}, is included as a potential application
of the spectral function analysis showing that strong spin-charge coupling does not
necessarily lead to superconductivity.

The {\it ab}-plane optical reflectance of optimally-doped Bi2212 single
crystals has been measured extensively \cite{quijada94,wang99}.  Furthermore, a
number of cobaltates Bi$_2$M$_3$Co$_2$O$_\delta$, (M=Ca, Sr, and Ba) have been
studied optically at room temperature \cite{watanabe91,terasaki93}. However,
with much improved signal-to-noise ratio, our detailed temperature-dependent
studies of optimally-doped Bi2212 and metallic
(Bi$_{0.5}$Pb$_{0.5}$)$_2$Ba$_3$Co$_2$O$_\delta$ have revealed a number of new
features.  For this study, large optimally-doped Bi2212 single crystals are
grown using the traveling-surface-floating-zone (TSFZ) method.  The single
crystal metallic cobaltate samples
(Bi$_{0.5}$Pb$_{0.5}$)$_2$Ba$_3$Co$_2$O$_\delta$ are prepared using a flux
technique \cite{loureiro01}.

The crystals are mounted on an optically-black cone, and the
temperature-dependent reflectance from 6~K to 295~K is measured in a
near-normal incidence arrangement from $\approx 100$ to over
$15\,000$~cm$^{-1}$ on a Bruker IFS 66v/S. The absolute reflectance is
determined by evaporating a gold film ($\approx 100$~nm in thickness) {\it in
situ} in a high vacuum ($\approx 1\times 10^{-8}$~Torr) over the sample at the
end of the reflectivity measurements at 295~K, and then measuring the
reflectance again \cite{homes93}.  The optical properties are determined from a
Kramers-Kronig analysis of the reflectance.  The conductivity data is analyzed
in the extended-Drude formalism \cite{puchkov96} with a frequency-dependent
scattering rate $\tau^{-1}(\omega)$ defined as
\begin{equation}
  \tau^{-1}(\omega)={ {\omega_p^2} \over {4\pi}} \Re \left[
  { {1} \over {\tilde\sigma(\omega)} } \right],
\end{equation}
where $\omega_p$ is the classical plasma frequency.

%
%
\begin{figure}[t]
\epsfxsize=5.5cm%
\vspace*{-0.2cm}%
\centerline{\epsffile{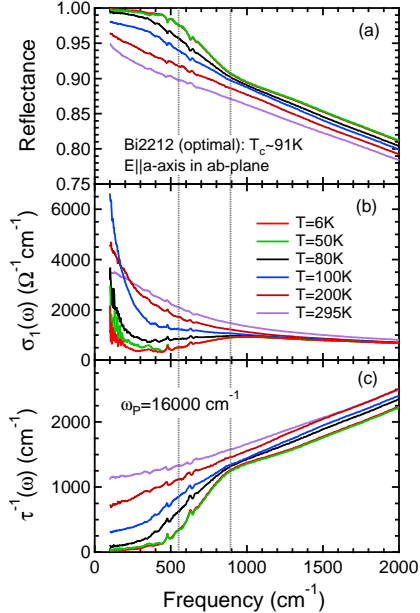}}%
\vspace*{-0.2cm}%
\medskip
\caption{The temperature-dependent {\it ab}-plane optical data of an
optimally-doped Bi2212 single crystal with the $E\parallel{a}$ from 100 to 2000
cm$^{-1}$.  (a) the temperature-dependent reflectance; (b) the
temperature-dependent $\sigma_1(\omega)$; (c) the temperature-dependent
$\tau^{-1}(\omega)$. }
\end{figure}

In Fig.~1, temperature-dependent {\it ab}-plane optical data are shown for an
optimally-doped Bi2212 single crystal with $E$-vector parallel to the {\it
a}-axis.  Two optic phonons at 477 and 630~cm$^{-1}$ have been observed for the
first time in conductivity measurements.  The temperature-dependent reflectance
is given in Fig.~1(a) from about 100 to 2000~cm$^{-1}$, $\sigma_1(\omega)$ in
Fig.~1(b), and the scattering rate $\tau^{-1}(\omega)$ calculated from Eq.~(1)
in Fig.~1(c). The value of $\omega_p = 16\,000$~cm$^{-1}$ is derived from the
optical conductivity sum rule at 295~K.  The temperature dependence of the
scattering rate $\tau^{-1}(\omega)$ is of interest here. Above 200~K, the
scattering rate $\tau^{-1}(\omega)$ is a monotonically increasing function with
frequency as shown in Fig.~1(c).  As the temperature approaches $T_c$ (about
91~K), abrupt curvature changes start to appear below 100~K, most noticeably as
a suppression of $\tau^{-1}(\omega)$ for frequencies less than about
600~cm$^{-1}$.  There are two positions where such changes in curvature can be
clearly identified at the lowest temperature, 6~K.  These two positions are
marked with two vertical lines in Fig.~1(c).  Similar behavior is observed in
the temperature-dependent reflectance in Fig.~1(a), and $\sigma_1(\omega)$ in
Fig.~1(b).

Carbotte {\it et al.} \cite{carbotte99} and Abanov {\it et al.}\cite{abanov01}
analyzed the features in $\tau^{-1}(\omega)$ by using an estimated spectral function:
\begin{equation}
  W(\omega)={1\over{2\pi}} {{d^2} \over {d\omega^2}} \left[
  {\omega\over{\tau(\omega)}} \right].
\end{equation}
In the case of BCS superconductors, $W(\omega)$ is closely related to the
electron-phonon spectral function $\alpha^2F(\omega)$
\cite{allen71,farnworth76,marsiglio98}, and Carbotte {\it et al.}
\cite{carbotte99} argued that for high $T_c$ cuprates, $W(\omega)$ can be
directly associated with the spin-charge excitation spectral density, $I^2
\chi^{\prime\prime}(\omega)$, derived from spin-polarized inelastic neutron
scattering. In particular, the peak in $W(\omega)$ is correlated to
$\Delta+\Delta_s$ below $T_c$.  In a more recent theoretical paper, Abanov {\it
et al.} \cite{abanov01} also studied $W(\omega)$ in cuprates by examining the
singularities in the optical response function. However, they argued that the
deep minimum in $W(\omega)$, which in their picture is located at
$2\Delta+\Delta_s$, is more relevant to superconductivity. They also identified
two weaker high frequency singularities at $4\Delta$ and $2\Delta+2\Delta_s$.

%
%
\begin{figure}[t]
\epsfxsize=5.5cm%
\vspace*{-0.2cm}%
\centerline{\epsffile{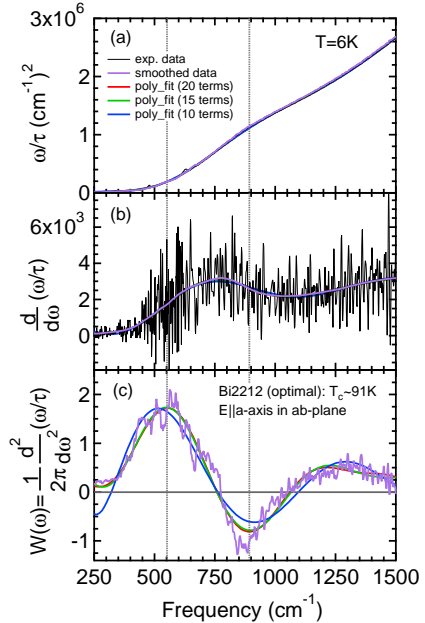}}%
\vspace*{-0.2cm}%
\medskip
\caption{An unambiguous method of extract the spectral function $W(\omega)$.
(a) The experimentally measured quantity $\omega/\tau$ at 6~K for
optimally-doped Bi2212, together with the smoothed experimental data and three
polynomial fits to the data; (b) the first derivatives of the five curves in
(a); (c) the second derivatives of the three polynomial fits to $\omega/\tau$,
and the second derivative of the smoothed $\omega/\tau$. }
\end{figure}

In Fig.~2 an unambiguous way of extracting the spectral function $W(\omega)$
from the experimental data is demonstrated.  The measured quantity
$\omega/\tau$ at 6~K is plotted for optimally-doped Bi2212 in Fig.~2(a),
together with the smoothed experimental data and three polynomial fits to the
data. In Fig.~2(b), the first derivatives of the curves shown in Fig.~2(a) are
presented.  The first derivative of the experimental data (the two optic
phonons at 477 and 630 cm$^{-1}$ are removed) is included to illustrate that it
has the same trend as the other curves.  In Fig.~2(c), the spectral function
$W(\omega)$ derived from different methods are plotted for $T=6$~K.  The
$W(\omega)$ determined from the smoothed $\omega/\tau$ is still very noisy, but
gives the same maximum and minimum as the polynomial fits. Because there are
too many different ways of smoothing the experimental data that do not always
give consistent results, fitting the experimentally measured quantity
$\omega/\tau$ with a high-order polynomial is adopted as the unambiguous method
of extracting the spectral function $W(\omega)$ from the optical data in this
study.

The temperature dependence of the spectral function $W(\omega)$ calculated
using the procedure outlined in the preceding paragraph is shown in Figs.~3(a)
and 3(b) for a optimally-doped Bi2212 single crystal and a untwinned
optimally-doped YBCO single crystal \cite{homes99}, respectively.   For the
optimally-doped Bi2212, there is a well defined maximum in $W(\omega)$ at
around 70~meV, a deep minimum at around 110~meV, and a weak high-energy maximum
at around 150~meV at 6~K. According the two theoretical pictures
\cite{carbotte99,abanov01} that rely on spin-charge coupling in high $T_c$
cuprates, these maxima and minima positions correspond to $\Delta+\Delta_s$,
$2\Delta+\Delta_s$ and $2\Delta+2\Delta_s$, respectively. One can therefore
uniquely determine the averaged values of the superconducting gap $\Delta=33\pm
3$~meV and the spin resonance $\Delta_s=41\pm 3$~meV from the optical data.
These optically obtained values agree very well with the directly measured
values of $\Delta=30$~meV \cite{valla00} and $\Delta_s=43$~meV \cite{fong99}
for optimally-doped Bi2212.  As the temperature increases to 80~K, the maximum
in $W(\omega)$ seems to split into two peaks while the position of the minimum
shows little change.  At a temperature of $T=100$~K, just above $T_c$, the
maximum in $W(\omega)$ has shifted to 40~meV which is precisely $\Delta_s$ for
optimally-doped Bi2212 \cite{fong99}, and the minimum has disappeared
completely into the noise.

%
%
\begin{figure}[t]
\epsfxsize=5.5cm%
\vspace*{-0.2cm}%
\centerline{\epsffile{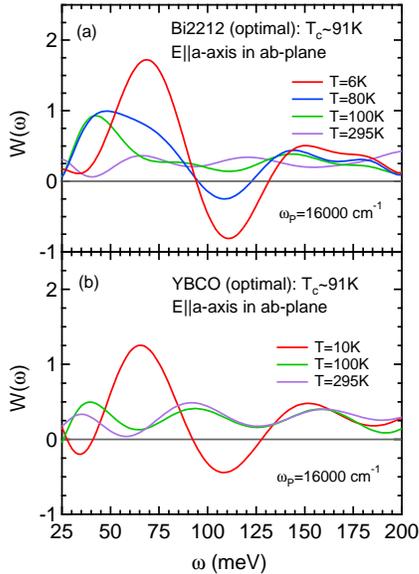}}%
\vspace*{-0.2cm}%
\medskip
\caption{The temperature dependence of the spectral function $W(\omega)$ of an
optimally-doped Bi2212 single crystal, and a untwinned optimally-doped YBCO
single crystal with $E\parallel{a}$ in the {\it ab}-plane (20 term polynomial
fits are used). (a) $W(\omega)$ of an optimally-doped Bi2212 single crystal at
four different temperatures; (b) $W(\omega)$ of a optimally-doped YBCO single
crystal at three different temperatures. }
\end{figure}

The temperature-dependent $W(\omega)$ for a untwinned optimally-doped YBCO
single crystal is shown in Fig.~3(b).  The general features of $W(\omega)$ at
the lowest temperature $T=10$~K is very similar to that of the optimally-doped
Bi2212, showing a maximum at around 67 meV, a minimum at around 105~meV and a
weaker maximum at around 145~meV.  One can again determine the averaged values
of the superconducting gap $\Delta=30 \pm 4$~meV and the spin resonance
$\Delta_s=40 \pm 4$~meV from the optical data.  These optically obtained values
also agree very well with the directly measured values of $\Delta=27$~meV
\cite{limonov00} and $\Delta_s=41$~meV \cite{bourges99} for optimally-doped
YBCO.  The main difference between the Bi2212 and YBCO data is the fact that
while there is a weak maximum at $\Delta_s$ in $W(\omega)$ above $T_c$ for
optimally-doped Bi2212 at 100~K, it is absent in optimally-doped YBCO
\cite{schachinger01}. This observation agrees with the suggestion that the
phase diagrams for Bi2212 and YBCO are different \cite{renner98}; namely there
is a pseudogap in optimally-doped Bi2212 above $T_c$, but there is no pseudogap
in optimally-doped YBCO.  Furthermore, the lineshape of the spectral function
at 80~K for optimally-doped Bi2212 indicates that the pseudogap and the
superconducting gap can coexist below $T_c$ adding support to the idea that the
two may have the same microscopic origin \cite{renner98}.  We have also made
some preliminary measurements on an La$_{2-x}$Sr$_x$CuO$_4$ ($x=0.17$) (LSCO)
sample ($T_c=38$~K) which does not show any prominent features in the spectral
function $W(\omega)$ that can be associated with superconductivity. This is
consistent with the neutron scattering results that do not show any prominent
spin resonance peaks in this system \cite{tranquada}. Therefore, the
theoretical picture of spin-charge coupling is in good agreement with the
optical data for all three high-$T_c$ cuprates.

%
%
\begin{figure}[t]
\epsfxsize=5.5cm%
\vspace*{-0.2cm}%
\centerline{\epsffile{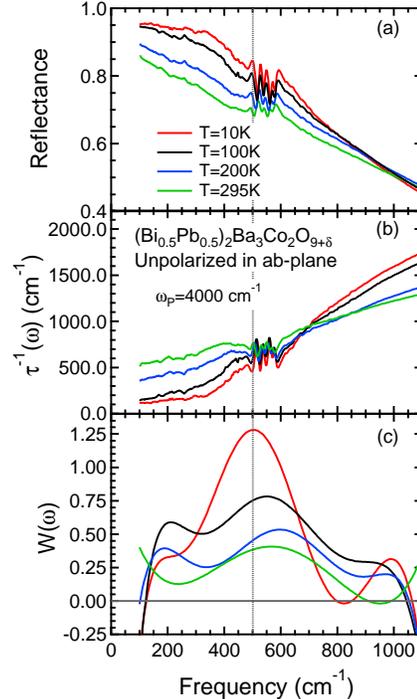}}%
\medskip
\vspace*{-0.2cm}%
\caption{The temperature-dependent unpolarized {\it ab}-plane optical data of a
(Bi$_{0.5}$Pb$_{0.5}$)$_2$\-Ba$_3$Co$_2$O$_\delta$ single crystal. (a) The
temperature-dependent reflectance; (b) the temperature-dependent
$\tau^{-1}(\omega)$; (c) the temperature-dependent spectral function
$W(\omega)$ (20 term polynomial fits are used).}
\end{figure}

As a potential application the spectral function $W(\omega)$ analysis, a
temperature-dependent optical study is carried out on a cobaltate,
(Bi$_{0.5}$Pb$_{0.5}$)$_2$Ba$_3$Co$_2$O$_\delta$.  No $T_c$ is observed and
this sample is shown to be metallic down to 30~mK \cite{loureiro01}.  The
temperature dependent optical data for
(Bi$_{0.5}$Pb$_{0.5}$)$_2$Ba$_3$Co$_2$O$_\delta$ is shown in Fig.~4. Because of
the low carrier density in this crystal, four sharp phonon peaks are observed
in reflectance and $\tau^{-1}(\omega)$ in the region between 450 and
650~cm$^{-1}$.  These phonon features are removed from the conductivity data
before fitting $\omega/\tau$ with polynomials.  The resulting temperature
dependent $W(\omega)$ is shown in Fig.~4(c). At the lowest temperature
$T=10$~K, there is a well defined maximum in $W(\omega)$ centered at around 500
cm$^{-1}$, or about 62~meV. This maximum is detectable to about 200~K.
However, there is no well defined minimum observed in $W(\omega)$ at any
temperature. While there is no superconductivity, these observations indicate
that there is also strong spin-charge coupling in this cobaltate.  If the
analysis of $W(\omega)$ is valid here, there should be a spin resonance peak at
around 60~meV below 200~K in this material and this should be verifiable by
inelastic neutron scattering.

In view of the temperature-dependent behavior of the spectral function
$W(\omega)$ of optimally-doped Bi2212, YBCO and metallic
(Bi$_{0.5}$Pb$_{0.5}$)$_2$Ba$_3$Co$_2$O$_\delta$ single crystals, it is clear
that both the maxima and minima in $W(\omega)$ should be examined at the same
time.  As one can see from our analysis of $W(\omega)$, the behavior of the
strong maximum in $W(\omega)$ is well described by the theory developed by
Carbotte {\it et al.} \cite{carbotte99}.  However, by strictly comparing
$W(\omega$) to $I^2\chi^{\prime\prime}(\omega)$ obtained from neutron
scattering, one ignores the negative part of $W(\omega)$ because
$I^2\chi^{\prime\prime}(\omega)$ is a positive quantity.  Thus the well defined
minimum in $W(\omega)$ is neglected.  In the more complete picture presented
recently by Abanov {\it et al.} \cite{abanov01}, both the maxima and the minima
in $W(\omega)$ are examined.  In general, their calculated spectral function
$W(\omega)$ agrees quite well with our experimentally extracted $W(\omega)$
except that there is no indication of the weak singularity at $4\Delta$ in the
experimental $W(\omega)$.  On the other hand, the physically intuitive
interpretation of the maximum in $W(\omega)$ as being directly related to
$I^2\chi^{\prime\prime}(\omega)$ is somewhat obscured. Further theoretical
study is underway \cite{carbotte} to bridge the gap between these two different
approaches.  A comment should be made here that even though our experimental
data and its analysis presented here are consistent with the theoretical
pictures of Carbotte {\it et al.} \cite{carbotte99} and Abanov {\it et al.}
\cite{abanov01}, the spectral function analysis alone does not exclude other
coupling mechanisms, e.g. charge-phonon coupling, and there is much evidence
that nanoscale charge inhomogeneities may also play an important role in these
systems.  One example of this is the stripe picture \cite{emery97}. In
particular, the lack of experimental evidence for strong spin-charge coupling
in LSCO systems poses some problems for spin-mediated superconductivity in high
$T_c$ cuprates \cite{carbotte99,abanov01} and the data on the metallic (but non-SC)
cobaltate indicates that strong spin-charge coupling maybe a necessary but not
exclusive condiction for high-$T_c$ superconductivity.

In conclusion, we have carried out temperature-dependent analysis of the
spectral function $W(\omega)$ for the optimally-doped
Bi$_2$Sr$_2$CaCu$_2$O$_{8+\delta}$, YBa$_2$Cu$_3$O$_{6.95}$ and the metallic
(Bi$_{0.5}$Pb$_{0.5}$)$_2$Ba$_3$Co$_2$O$_\delta$ single crystals. The behavior
of the maxima and minima in $W(\omega$) is consistent with the picture of
spin-charge coupling in high-$T_c$ cuprates. More experimental studies in the
under-doped and over-doped regions of Bi2212 as well as in the metallic and
insulating regions of cobaltates are now underway to further our understanding
of these systems.

%
%
We would like to thank J.P. Carbotte, G.L. Carr, A.V. Chubukov, V.J. Emery,
P.D. Johnson, A. Millis, T. Timusk, T. Valla, and Z. Yusof for helpful
discussions. The work was supported by the U.S. Department of Energy under
Contract No. DE-AC02-98CH10886. Research undertaken at NSLS was supported by
the U.S. DOE, Division of Materials and Chemical Sciences.

\vspace*{-0.5cm}

%
%
%

%
%
%

\begin{references}
%
\vspace*{-1.5cm}
%
\bibitem[*]{correspond} Electronic address: myron@bnl.gov
%
\bibitem{carbotte00} J.P.~Carbotte, Physics in Canada {\bf 56}, 257
 (2000).
%
\bibitem{mcmillian65} W.L.~McMillian and J. M. Rowell, Phys. Rev. Lett.
 {\bf 14}, 108 (1965).
%
 \bibitem{allen71} P.B.~Allen, Phys. Rev. B {\bf 3}, 305 (1971).
%
\bibitem{farnworth76} B.~Farnworth and T. Timusk, Phys. Rev. B {\bf 10},
 5119 (1976).
%
\bibitem{marsiglio98} F.~Marsiglio, T. Startseva, and J. P. Carbotte,
 Phys. Lett. A {\bf 245}, 172 (1998).
%
\bibitem{carbotte99} J.P.~Carbotte, E. Schachinger, and D.N. Basov,
 Nature (London) {\bf 401}, 354 (1999).
%
\bibitem{abanov01} A.~Abanov, A. V. Chubukov, and J. Schmalian,
 cond-mat/0101220, (2001).
%
\bibitem{watanabe91} Y.~Watanabe {\it et al.}, Phys. Rev. B {\bf 43}, 3026 (1991).
%
\bibitem{terasaki93} I.~Terasaki, T. Nakahashi, A. Maeda, and K. Uchinokura,
 Phys. Rev. B {\bf 47}, 451 (1993).
%
\bibitem{loureiro01} S.M.~Loureiro {\it et al.}, Phys. Rev. B {\bf 63}, 094109
 (2001).
%
\bibitem{quijada94} M.A.~Quijada {\it et al.}, Physica C {\bf 235-240},
 $14\,1123$ (1994); A.V.~Puchkov {\it et al.}, Phys. Rev. Lett. {\bf 77}, 3212
 (1996); M.A.~Quijada {\it et al.}, Phys. Rev. B {\bf 60},
 $14\,917$ (1999).
%
\bibitem{wang99} N.L.~Wang, A. W. McConnell, and B. P. Clayman, Phys.
 Rev. B {\bf 59}, 576 (1999).
%
\bibitem{homes93} C.C.~Homes, M. Reedyk, D. Crandles, and T. Timusk,
Appl. Opt. {\bf 32}, 2972 (1993).
%
\bibitem{puchkov96} A.V.~Puchkov, D. N. Basov, and T. Timusk, J. Phys. C
 {\bf 8}, $10\,049$ (1996).
%
\bibitem{homes99} C.C. Homes {\it et al.}, Phys. Rev. B {\bf 60}, 9782 (1999).
%
\bibitem{valla00} T.~Valla {\it et al.}, Phys. Rev. Lett. {\bf 85}, 828
 (2000).
%
\bibitem{fong99} H.F.~Fong {\it et al.}, Nature (London) {\bf 398}, 588
 (1999).
%
\bibitem{limonov00} M.F.~Limonov, A.I. Rykov, S. Tajima, and A.
 Yamanaka, Phys. Rev. B {\bf 61}, $12\,412$ (2000).
%
\bibitem{bourges99} P.~Bourges {\it et al.}, in High Temperature
 Superconductors, edited by S. E. Barnes (American Institute of Physics,
 Amsterdam, 1999), p. 202.
%
\bibitem{schachinger01} E.~Schachinger, J.P. Carbotte, and D.N. Basov,
 Europhys. Lett. (in press).
%
\bibitem{renner98} Ch.~Renner {\it et al.}, Phys. Rev. Lett. {\bf 80}, 149
 (1998); \O. Fischer, Bull. Amer. Phys. Soc. {\bf 46}, 925 (2001).
%
\bibitem{tranquada} J.M.~Tranquada (private communication).
%
\bibitem{carbotte} J.P.~Carbotte (private communication).
%
\bibitem{emery97} V.J.~Emery, S.A. Kivelson, and O. Zachar, Phys. Rev. B
 {\bf 56}, 6120 (1997).
%
\end{references}
\end{document}